\documentstyle[preprint,eqsecnum,aps]{revtex}
\tightenlines
\begin{document}

\draft \title{\Large A Renormalization Group Study of the
  $(\phi^*\phi)^3$ Model coupled to a Chern-Simons
  Field\footnote{Copyright 1999 by The American Physical Society}}
\author{V. S. Alves*, M. Gomes, S. L. V. Pinheiro* and A. J. da Silva}
\address{Instituto de F\'\i sica, USP
  C.P. 66318 - 05315-970, S\~ao Paulo - SP, Brazil\\
  **Departamento de F\'\i sica, Universidade Federal do Par\'{a} -
  Bel\'{e}m -PA,Brazil}

\date{October 7, 1999} 

\maketitle

\begin{abstract}
  We consider the model of a massless charged scalar field, in $(2+1)$
  dimensions, with a self interaction of the form $\lambda (\phi^*
  \phi)^3$ and interacting with a Chern Simons field. We calculate the
  renormalization group $\beta$ functions of the coupling constants
  and the anomalous dimensions $\gamma$ of the basic fields. We show
  that the interaction with the Chern Simons field implies in a
  $\beta_{\lambda}$ which suggests that a dynamical symmetry breakdown
  occurs.  We also study the effect of the Chern Simons field on the
  anomalous dimensions of the composite operators $(\phi^* \phi)^n$,
  getting the result that their operator dimensions are lowered.
\end{abstract}
\vspace{5cm}

\newpage
\section{Introduction}
Self interacting scalar fields are the simplest nontrivial field
theories. Nevertheless they have found large application in many
different phenomena. Renormalization group analyses of the model of
scalar fields in $(2+1)$ dimensions, with a self interaction of the
form $\lambda \phi^6$ have appeared in the literature \cite{1} in
conjunction with other self interactions, and also in interaction with
other fields. On the other side, the Chern-Simons (CS) field theory
\cite{2} is known to cause some strange effects in matter fields, the
most known being the transmutation of their spins and statistics
\cite{3}.
 
Bosons (fermions) interacting with a CS field get an extra
contribution to their spins and statistical phases, changing to anyons
and even to fermions (bosons). Studies of the change in the scale
behavior of matter fields due to their interaction with the CS field
have also been considered \cite{4,5,6}.

In this paper we study the model of a massless charged scalar field
with a self interaction of the form $\lambda (\phi^{*} \phi)^3$ and
interacting with an Abelian CS field. Classically it only involves
dimensionless parameters and is scale invariant. It is also strictly
renormalizable: no induction of terms of the forms $m^2 (\phi^* \phi)$
or $g (\phi^* \phi)^2$ occurs. Besides the calculation of the
anomalous dimensions of $\phi$ and $A^\mu$ and the $\beta$ functions
related to their coupling constants, we also calculate the anomalous
dimensions of composite operators of the form $(\phi^* \phi)^n$.
Some of our conclusions agree and others disagree with the previous
literature. This will be discussed in section III and in the Conclusions.

To regulate the ultraviolet (UV) behavior we use a simplified version
of dimensional regularization, the so called ``Dimensional Reduction''
method. It consists of contracting and simplifying the Lorentz
tensors, before extending the Feynman integrals out of 3 dimensions.
This procedure, previously used by several authors \cite{4,5,6},
greatly simplifies calculations involving the CS field, because it
does not require the extension of the Levi-Civita tensor
$\varepsilon^{\mu\nu\rho}$ out of $3$ dimensions.  In Feynman
integrals only involving scalar vertices and propagators, no
difference appear between the results gotten by using one or the other
method. In graphs involving the CS field and the
$\varepsilon^{\mu\nu\rho}$, the differences of this method to a
``full'' dimensional regularization would only show up \cite{4} in
sub-leading contributions to the Feynman integrals; that is, if $D$
stands for the extended dimension of the space time when the Feynman
integrals are expanded in Laurent series in $\epsilon\equiv(D-3)$, no
difference in the leading divergent term in $1/\epsilon$ will appear.
It is, on the other side, a characteristic of dimensional
regularization in $(2+1)$ dimensions, that one loop graphs are finite,
and 2 loops graphs have at most a single pole divergence in
$\epsilon$. As the calculation of the renormalization group parameters
only involve the use of the divergent parts of the graphs, no
differences to the full dimensional regularization is expected up to 2
loops in graphs that involve the CS propagator, and any number of
loops in graphs only involving the scalar propagator. In this paper we
will restrict the calculations to up 2 loops in all graphs involving
propagators of the CS field, and 4 loops in graphs involving only the
propagator of the scalar field.  As we will explicitly show that ( at
least ) to that orders, dimensional reduction is enough to regularize
the model and to preserve the gauge symmetry, as expressed by the Ward
Identities (WI).  We will work in the Landau gauge and, avoiding
exceptional momenta, no infrared (IR) divergences appear.

The plan of the paper is as follows. In section II the model is
presented, and the divergent UV counterterms, necessary for the
renormalization group study, are obtained by calculating the CS 2
point function and the scalar field 2 and 6 point functions. In
section III the renormalization group $\beta$ functions and anomalous
dimensions of the fields are obtained, and compared with other
calculations. The change in the dynamical behavior of the $\phi$ field
due to the interaction of the CS field is discussed. The influence of
the CS in the dimension and renormalizability of operators of the form
$(\phi^*\phi)^n$ is also studied. A summary of the results are
presented in the Conclusions.  In Appendix A the explicit verification
of the WI is given, and in Appendix B some Feynman integrals are
calculated as examples.

\section{The Model}

The model is constituted by a massless charged boson in $2+1$ dimensions 
represented by a field $\phi$, with a self interaction of the form
 $(\phi^*\phi)^3$  and minimally interacting with a Chern-Simons (CS)
 field $A_\mu$. Its Lagrangian density is given by 
\begin{eqnarray}
{\cal L} &=& \partial_\mu \phi_0^\dagger \partial^\mu \phi_0
- ie_0 A^\mu_0 (\phi_0^\dagger \partial_\mu \phi_0 - \partial_\mu
\phi_0^\dagger \phi_0) 
+ e_0^2 A^\mu_0 A_{0\mu} (\phi_0^\dagger\phi_0)
\nonumber \\
&-& \frac{\lambda_0}{6^2} (\phi_0^\dagger\phi_0)^3 + 
\frac12 \varepsilon_{\mu\nu\rho} A^\mu_0 \partial^\nu A^\rho_0.   
\label{2.1}
\end{eqnarray}
The metric is $g_{\mu\nu}=(1,-1,-1)$, $\partial_\mu$ stands for
$\frac{\partial}{\partial x^\mu}$, $\varepsilon_{\mu\nu\rho}$ is the
antisymmetric Levi-Civita tensor with $\varepsilon^{012}=1$, and $e_0$
and $\lambda_0$ are dimensionless coupling constants.
The subscript ``$0$'' means that the corresponding quantity is 
``unrenormalized''. 

The model is renormalizable: all the UV infinities of the perturbative
series can be absorbed in a redefinition of the unrenormalized
quantities. It also has a gauge symmetry what suggests the use of
dimensional regularization \cite{8}. However, the presence of the
Levi-Civita tensor in the CS term makes dimensional regularization
cumbersome and the calculations become awkward in more than one loop.
We will so, take advantage of some characteristics of $(2+1)$
dimensions and use a simplified version of Dimensional Regularization,
the so called Dimensional Reduction\cite{4,5}. In this procedure, the
Lorentz tensor algebra is considered in $(2+1)$ dimensions and only
the remaining scalar Feynman integrals are extended out of $(2+1)$
dimensions. It was verified in \cite{5}, up to 2 loops, that for the
non-Abelian Chern-Simons theory, this procedure, in fact preserves the
Slavnov-Taylor identities.  As we will also show below up to 2 loops,
it also preserves the Ward identities in our model, and no
inconsistencies appear.

To get information on the asymptotic behavior of the model, we need to
calculate the renormalization group parameters: $\beta$ functions and
anomalous dimensions of the fields.  For this task, adopting the
Renormalization Group approach of t'Hooft\cite{9} based on minimal
subtraction, we only need to calculate the divergent parts of some
vertex functions, more precisely the residues of the poles in
$1/\epsilon$, where $\epsilon=3-D$ and $D$ is the
``extended''dimension of the space time. In $(2+1)$ dimension this
means that we must go to at least 2 loops calculations, because as a
characteristic of dimensional regularization, one loop integrals are
finite.

Introducing the renormalized fields $\phi$ and $A_\mu$ and the
renormalized coupling constants $e$ and $\lambda$ through the
definitions
\begin{equation}
\phi_0 = Z^{\frac{1}{2}}_\phi \phi = (1+A)^{\frac{1}{2}} \phi
\label{2.2a}
\end{equation}
\begin{equation}
A^\mu_0 = Z^{\frac{1}{2}}_A A^\mu = (1+B)^{\frac{1}{2}} A^\mu
\label{2.2b}
\end{equation}
\begin{equation}
e_0 = e\mu^{\frac{\epsilon}{2}}  (1+D)/Z_\phi Z_A^{\frac{1}{2}} 
\label{2.2c}
\end{equation}
\begin{equation}
e_0^2 =e^2 \mu^{\epsilon} (1+E)/Z_\phi Z_A 
\label{2.2d}
\end{equation}
\begin{equation}
\lambda_0 =\mu^{2\epsilon} (\lambda+C)/Z_\phi^3  
\label{2.2e}
\end{equation}
where $\mu$ is a mass parameter introduced to keep $e$ and $\lambda$
dimensionless quantities, and $A$ to $E$ are the counterterms to be
chosen so as to make the renormalized quantities finite, in each order
of perturbation. As will be seen in the calculations, the
renormalization of $\lambda$ in presence of the CS field is not
multiplicative. By substituting these definitions in (\ref{2.1}) we
get for ${\cal L}$:
\begin{eqnarray}
{\cal L} &=& \partial_\mu \phi^\dagger \partial^\mu \phi+
\frac{1}{2} \varepsilon_{\mu\nu\rho} A^\mu \partial^\nu A^\rho  
- ie\mu^{\frac{\epsilon}{2}} A^\mu (\phi^\dagger \partial_\mu \phi 
-\partial_\mu \phi^\dagger \phi) 
+ e^2\mu^\epsilon A^\mu A_{\mu} (\phi^\dagger\phi)
\nonumber \\
&&- \frac{\lambda \mu^{2\epsilon}}{6^2} (\phi^\dagger\phi)^3  
 + A\partial_\mu \phi^\dagger \partial^\mu \phi
+\frac{B}{2} \varepsilon_{\mu\nu\rho} A^\mu \partial^\nu A^\rho
- ie\mu^{\frac{\epsilon}{2}} D A^\mu (\phi^\dagger \partial_\mu \phi 
- \partial_\mu \phi^\dagger \phi) 
\nonumber \\
& &
+ e^2\mu^\epsilon E A^\mu A_{\mu} (\phi^\dagger\phi)
- \frac{\mu^{2\epsilon}C}{6^2} (\phi^\dagger\phi)^3 \; .
\label{2.3}
\end{eqnarray}

The Feynman rules for this Lagrangian in the Landau gauge are depicted
in figure 1. This gauge can be implemented by adding to the Lagrangian
a gauge fixing term: $\frac{1}{2\xi} (\partial^\mu A_\mu)^2$,
inverting the free quadratic part of the $A^\mu$ to get the CS
propagator and, then letting $\xi \rightarrow \infty$. The would be
Faddeev-Popov ghost field is completely decoupled of the other fields
and does not have any effect. Calling $\Gamma (p)$ the scalar field
1PI two point function, and $\Gamma^\mu (q;p,p^\prime)$ and
$\Gamma^{\mu\nu} (q,k;p,p^\prime)$, respectively the trilinear and
quadrilinear CS scalar field vertices, where $q$ and $k$ represent
``photon'' momenta and $p$ and $p^\prime$ scalar field momenta, we
have the WI
\begin{equation}
q^\mu \Gamma_\mu (q;p,p^\prime) = -e [\;\Gamma
(p^\prime)- \Gamma (p)\;]
\label{2.4a}
\end{equation}

\begin{equation}
q^\mu \Gamma_{\mu\nu} (q,k;p,p^\prime) = -e
 [\;\Gamma_\nu (k;p+q,p^\prime)- \Gamma_\mu (k;p,p^\prime-q)\;]\; ,
\label{2.4b}
\end{equation}
which require that $E=D=A$, leaving us with only three (we choose $A$,
$B$, $C$) counterterms to be fixed. An explicit proof of these WI in
two loops is given in the Appendix A.

To determine $A$, $B$, and $C$ we need to calculate the simple pole
part of the 2 point function of the CS field, $\Pi_{\mu\nu} (q)$), and
the scalar field 2 and 6 point functions, respectively $\Gamma_2$ and
$\Gamma_6$.  In graphs involving the CS field, we will extend the
calculations up to two loops getting at most a simple pole in
$1/\epsilon$; in graphs only involving the scalar field we will go up
to four loops.  So, in the tensorial Feynman integrals, in which
dimensional reduction could possibly differ from dimensional
regularization ( in the sub leading terms in $1/\epsilon$ ) no
difference between the two methods are expected in the calculation of
the counterterms and in the renormalization group parameters.

Let us start with $\Pi_{\mu\nu}$. The only divergent diagrams, up to 2
loops, are those shown in figure 2 ( the possible counterterm is also
drawn in the figure ).  Their contributions are given by
\begin{equation}
(2a) = 4e^4 \int {\cal D} q \int {\cal D} k
\frac{\varepsilon_{\mu\nu\rho}k^\rho}{k^2 (q-p)^2 (q+k)^2}
\label{2.5a}
\end{equation}
\begin{equation}
(2b) = e^4 \int {\cal D} q \int {\cal D} k
\frac{(2k+q)^\alpha \varepsilon_{\alpha\beta\gamma}q^\gamma
  (2k+q-2p)^\beta (2k-p)_\mu (2k+2q-p)_\nu}
{k^2 (k+q)^2 q^2 (k-p)^2 (q+k-p)^2}\; ,
\label{2.5b}
\end{equation}
where ${\cal D} q \equiv \mu^\epsilon d^{3-\epsilon}
q/(2\pi)^{3-\epsilon}$ and an infinitesimal imaginary part is supposed
in every propagator denominator ($p^2 \rightarrow p^2+i\eta$, $\eta\ll
1$). Both integrals are logarithmically divergent. The divergent parts
are of the form $\varepsilon_{\mu\nu\rho} p^\rho I$ where $I$ is a
scalar integral, that can be calculated by the usual dimensional
continuation, after reducing the denominator to a single monomial
through the use of Feynman parameters. The results are
\begin{equation}
(2a) = 4 \varepsilon_{\mu\nu\rho} p^\rho
\left(-\frac{e^4}{96\pi^2}\frac{1}{\epsilon}\right) 
+ \mbox{finite part}
\label{2.6a}
\end{equation} 
\begin{equation}
(2b) = \varepsilon_{\mu\nu\rho} p^\rho
\left(\frac{e^4}{24\pi^2}\frac{1}{\epsilon}\right) 
+ \mbox{finite part}\; .
\label{2.6b}
\end{equation}
As can be seen, the divergent parts of the two integrals cancel each
other and we are left with only finite contributions to
$\Pi_{\mu\nu}$. So, the counterterm $B$ can be chosen as $B=0$, and no
infinite wave function renormalization of the CS field interacting
with massless scalar field is needed. This result extends for massless
matter, the result of the Coleman-Hill theorem \cite{10}.

Let us now look at the scalar two point function, $\Gamma_2 (p)$. The
divergent graphs up to second order in $\alpha$ and $\lambda$ are
shown in figure 3, together with the counterterm. Their contributions
are given by
\begin{equation}
(3a)= -2 e^4 i \int {\cal D} q \frac{1}{(p+q)^2} \int {\cal D} k 
\;\varepsilon_{\mu\nu\rho} \frac{k^\rho}{k^2}\; \varepsilon^{\nu\mu\gamma}
\frac{(k+q)_\gamma}{(k+q)^2} 
\label{2.7a}
\end{equation}
\begin{equation}
(3b)=  e^4 i^3 \int {\cal D} q \;\varepsilon_{\alpha\beta\gamma}
\frac{q^\gamma}{q^2} \frac{(2p+q)^\beta}{(p+q)^2}
\int {\cal D} k \frac{(2p+k)^\mu (2p+2q+k)^\nu (2k+2p+q)^\alpha}
{(p+k)^2(p+q+k)^2} 
\;\varepsilon_{\mu\nu\rho} \frac{k^\rho}{k^2}  
\label{2.7b}
\end{equation}

\begin{equation}
(3c)=  e^4 i^3 \int {\cal D} q \frac{1}{(p+q)^2}\;
\varepsilon_{\mu\alpha\lambda}
\frac{q^\lambda}{q^2} \;\varepsilon_{\beta\nu\rho} \frac{q^\rho}{q^2}
\int {\cal D} k \frac{(2k+q)^\alpha (2k+q)^\beta (2p+q)^\mu (2p+q)^\nu}
{k^2(k+q)^2}   
\label{2.7c}
\end{equation}
\begin{equation}
(3d)= -\frac{\lambda^2}{2^2.3} i^5 \int {\cal D} k_1 {\cal D} k_2
{\cal D} k_3 {\cal D} k_4 \frac{1}{k_1^2 k_2^2 k_3^2 k_4^2 (p+k_1 +k_2
  -k_3 -k_4)^2}\; .
\label{2.7d}
\end{equation}

After the simplification of the tensor algebra in $(2+1)$ dimensions
we are left with multiple scalar integrals that can be made, one loop
at time, through the reduction of the denominators by successive use of
Feynman parameters. The results are
\begin{equation}
(3a)=-2i e^4  \left( \frac{p^2}{96\pi^2} \frac{1}{\epsilon} + ...\right)
\label{2.8a}
\end{equation}
\begin{equation}
(3b)=-i e^4  \left( \frac{p^2}{12\pi^2} \frac{1}{\epsilon} + ...\right)
\label{2.8b}
\end{equation}
\begin{equation}
(3c)=-i e^4  \left( \frac{p^2}{24\pi^2} \frac{1}{\epsilon} + ...\right)
\label{2.8c}
\end{equation}
\begin{equation}
(3d)=- i\frac{\lambda^2}{2^2.3} \left( -\frac{p^2}{3.2^{11}\pi^4} 
\frac{1}{\epsilon} + ...\right)\; .
\label{2.8d}
\end{equation}

For the contribution $(iA p^2)$ of the counterterm to cancel these
divergences we must choose:
\begin{equation}
A= \left( \frac{7}{48\pi^2} \alpha^2 -\frac{1}{3^2.2^{13} \pi^4}
\lambda^2 \right) \frac{1}{\epsilon} \; .
\label{2.9}
\end{equation}

Let us now proceed to the calculation of $C$, the counterterm of the
coupling constant $\lambda$. For this task we need to get the
divergent parts of $\Gamma_6 (p_1,...,p_6)$. After a lengthly analyses
of the many graphs involved, we are left with the divergent
contributions drawn in figure 4. The bullets on the diagrams $4p, 4q,
4r$ $4s$ and $4t$ mean the insertion of the counterterm in the
corresponding vertex.  The calculation of all these diagrams can be
reduced to the calculation of the nine integrals represented in
figure 5. In the Appendix B we show, as examples, the calculation (of
the divergent parts) of $5a, 5b, 5d$ and $5f$. Here we present only
the results:
\begin{equation}
{\cal G} (p,q) = -\frac{1}{2^5 \pi^2} \frac{1}{\epsilon} +
\mbox{finite part},
\label{2.10a}
\end{equation}
\begin{equation}
{\cal H} (p) = \frac{i}{16 \pi^2} \frac{1}{\epsilon} +
\mbox{finite part},
\label{2.10b}
\end{equation}
\begin{equation}
{\Delta_3} (p) = -\frac{i}{2^5 \pi^2} \left[\frac{1}{\epsilon} +
+\left( \log \frac{4\pi \mu^2}{-p^2} - 3-2 \gamma -2 \log 2 \right) +
{\cal O} (\epsilon) \right],
\label{2.10c}
\end{equation}
\begin{equation}
{\cal Y} (p) = -\frac{1}{2^{12} \pi^4} \frac{1}{\epsilon} +
\mbox{finite part},
\label{2.10d}
\end{equation}
\begin{equation}
{\cal Z} (p,q) = \frac{1}{2^{11} \pi^4} \frac{1}{\epsilon} +
\mbox{finite part},
\label{2.10e}
\end{equation}
\begin{equation}
{\cal W} (q,p) = -\frac{1}{2^{11} \pi^4} \left[\frac{1}{\epsilon^2} +
\frac{1}{\epsilon}\left( 2\log \frac{4\pi \mu^2}{-(p+q)^2}
  +8-\frac{11}{2} \gamma  \right) +\mbox{finite part}\right],\\
\label{2.10f}
\end{equation}
\begin{equation}
{\cal M} (a,c,d) = \frac{3 i}{2^6 \pi^2} \frac{1}{\epsilon} 
+ \mbox{finite part},
\label{2.10g}
\end{equation}
\begin{equation}
{\cal N} (a,c,d) = \frac{i}{2^5 \pi^2} \frac{1}{\epsilon} 
+\mbox{finite part},
\label{2.10h}
\end{equation}
and
\begin{equation}
{\cal Q}(a,b,c)=\frac1{2^5\pi^2}\frac1\epsilon+\mbox{finite part}
\end{equation}
where $\gamma$ is the Euler constant. In some graphs we will need 
the result of ${\Delta_3}^2 (p)$:
\vspace{.5cm}
\begin{equation}
{\Delta_3}^2 (p) = -\frac{1}{2^{10}\pi^4} \left[ \frac{1}{\epsilon^2}
+  \frac{1}{\epsilon} \left( 2\log \frac{4\pi \mu^2}{-p^2} 
-2(3+2\gamma + 2 \log 2) \right)+ \mbox{finite part}\right]\; .\\
\nonumber
\end{equation}
By collecting all contributions of figure 4 we can write
\vspace{.5cm}
\begin{eqnarray}
\Gamma_6 (p_1,&p_2&,p_3,p_1^\prime,p_2^\prime,p_3^\prime)
\,\,\mu^{-2\epsilon}\nonumber\\ 
&=& -\, \frac{\lambda^2}{6} {\Delta_3}(p_1+p_2+p_3)
-  \frac{\lambda^2}{2}[ {\Delta_3}(p_1+p_2-p_2^\prime) + 8
\;\mbox{terms}]
\nonumber \\
& &+ 2\, i \lambda \alpha^2 [ {\cal G} (p_1,-p_1^\prime) + 8\; \mbox{terms}]
+ 2\, i \lambda \alpha^2 [ {\cal G} (p_1,p_2) + 2\; \mbox{terms}]
\nonumber \\
& & + 2\, i \lambda \alpha^2 [ {\cal G} (-p_1^\prime,-p_2^\prime) + 2\;
\mbox{terms}]
-2 \lambda \alpha^2 [ {\cal H} (p_1-p_1^\prime) + 8\; \mbox{terms}]
\nonumber \\
& &+\, i \frac{5}{4}\lambda^3 [  {\cal Y} (p_1-p_1^\prime,p_2-p_2^\prime) 
+ 5\; \mbox{terms}]
\nonumber \\
& &+\, i \frac{3}{4}\lambda^3 [  {\cal Y} (p_1+p_2,-(p_1+p_2^\prime)) 
+ 8\; \mbox{terms}]
\nonumber \\
& &+\, i \frac{1}{4}\lambda^3 [  {\cal Z} (p_1,p_2) +2\; \mbox{terms}]
-\frac{5}{12}\lambda^2 [  {\cal Z} (p_1,-p_1^\prime) +8\;
\mbox{terms}]
\nonumber \\
& &+\, i \frac{1}{4}\lambda^3 [  {\cal Z} (-p_1^\prime,-p_2^\prime) +2\;
\mbox{terms}]
\nonumber \\
& &+\, i \frac{1}{36}\lambda^3 [  {\Delta_3}^2 (p_1+p_2+p_3)]
+ i \frac{1}{4}\lambda^3 [  {\Delta_3}^2 (p_1+p_2-p_1^\prime) +8\;
\mbox{terms}]
\nonumber \\
& &+\, i \frac{1}{4}\lambda^3 [  {\cal W} (p_1,p_2+p_3) +2\;\mbox{terms}]
\nonumber \\
& &+\, i \frac{1}{4}\lambda^3 [  {\cal W}
(-p_1^\prime,-(p_2^\prime+p_3^\prime)) +2\;\mbox{terms}]
\nonumber \\
& &+\, i \frac{3}{4}\lambda^3 [  {\cal W}
(p_1,p_2-p_1^\prime) +17\;\mbox{terms}]
\nonumber \\
& &+\, i \frac{3}{4}\lambda^3 [  {\cal W}
(-p_1^\prime,p_1-p_2^\prime) +17\;\mbox{terms}]
\nonumber \\
& &+\, i \frac{7}{12}\lambda^3 [  {\cal W}
(p_1,-(p_1^\prime+p_2^\prime)) +8\;\mbox{terms}]
\nonumber \\
& &+\, i \frac{7}{12}\lambda^3 [  {\cal W}
(-p_1^\prime,p_1+p_2) +8\;\mbox{terms}]
-\frac{\lambda C}{3}  {\Delta_3} (p_1+p_2+p_3)
\nonumber \\
& &-\, \lambda C  [{\Delta_3} (p_1+p_2-p_1^\prime) +8\;\mbox{terms}]
- i C\nonumber\\
& &+\, 2^4 \alpha^4 [ {\cal M} (p_1,p_2-p_2^\prime,p_3-p_3^\prime) +
17 \mbox{terms}] \nonumber\\
& &+\, 2^5 \alpha^4 [ {\cal N} (p_1,p_2-p_2^\prime,p_3-p_3^\prime) +
17 \mbox{terms}] \,\nonumber\\
&& + i 2^2\alpha^4[{\cal Q}(p_1-p_{1}^{\prime},p_2-p_{2}^{\prime},p_3-p_{3}^{\prime})+35 \mbox{terms}]
\label{2.12}
\end{eqnarray}
from which, after imposing that the result be finite, we get
\begin{eqnarray}
C =& & \lambda^2 \frac{7}{48 \pi^2} \frac{1}{\epsilon}
-\lambda \alpha^2 \left[\frac{33}{16\pi^2}\right] \frac{1}{\epsilon} 
+ \alpha^4 \frac{72}{2 \pi^2} \frac{1}{\epsilon}.\nonumber\\[0.5cm]
& &- \lambda^3 \left[ \frac{582 + 57 \pi^2 -1092\gamma}{2^{14} \pi^4} \right]
\frac{1}{\epsilon} 
+ \lambda^3 \left[ \frac{49}{2^8 3^2 \pi^4} \right] \frac{1}{\epsilon^2},
\label{2.13}
\end{eqnarray}
The term proportional to
$\alpha^4$ in the above expression shows that the renormalization of
$\lambda$ is not multiplicative, a fact that will lead to an
interesting effect in the renormalization group equations. In the next
section, results (\ref{2.9}), (\ref{2.13}) and $B=0$ will be used to
determine the renormalization group parameters.

%%%%%%%%%%%%%%%%%%%%%%%%%%%%%%%%%%%%%%%%%%%%%%%%%%%%%%%%%%%%%%%%%%%%%%

\section{Renormalization Group Analyses}

Let us start by verifying the that the CS coupling does
not run. Equation (\ref{2.2d}) is
\begin{equation}
\alpha_0 = \alpha \mu^\epsilon \frac{(1+E)}{(1+A) (1+B)}\; .
\label{3.1}
\end{equation}
As we have seen in the last section, $B=0$ and, as consequence of the
Ward Identities, we also have $E=A$. Thus (\ref{3.1}) reduces to

\begin{equation}
\alpha_0 = \alpha \mu^\epsilon \; ,
\label{3.2}
\end{equation}
from which, in the way of \cite{9} we get
\begin{equation}
0\equiv \mu^{1-\epsilon} \frac{d\alpha_0}{d\mu} = \epsilon \alpha +
\mu \frac{d\alpha}{d\mu}\;,
\label{3.3}
\end{equation}
and therefore 
\begin{equation}
\beta_{\alpha}
 = \mu \left.\frac{d\alpha}{d\mu}\right|_{\epsilon\rightarrow 0}
\rightarrow 0 \; ,
\label{3.4}
\end{equation}
showing that $\alpha$ does not run under a rescaling of $\mu$ or the
momenta of the Green function. A similar result was get in \cite{5} 
for a model of a scalar field interacting with a non Abelian CS field.
These results extend to massless matter, the result of the theorem 
of Coleman-Hill \cite{10}.  

For calculating $\beta_{\lambda}$ we start with equation (\ref{2.2e}):
\begin{equation}
\lambda_0 = \mu^{2\epsilon} \frac{\lambda + C }{(1+ A )^3}
= \mu^{2\epsilon} (\lambda + C - 3 A + \cdots) 
\label{3.5}
\end{equation}
By substituting (\ref{2.9}) and (\ref{2.13}) in this
equation we get
\begin{equation}
\lambda_0 = \mu^{2\epsilon} \left(\lambda 
+\frac{\lambda_1 (\alpha,\lambda)}{\epsilon} + \cdots \right)\; ,
\label{3.7}
\end{equation}
where
\begin{equation}
\lambda_1 (\alpha,\lambda) = a (\lambda^2 - c \alpha^2\lambda
+ d \alpha^4  - b\lambda^3 )\;
\label{3.8}
\end{equation}
with
\begin{equation}
a= \frac{7}{48\pi^2} = 0.01478,
\label{3.9a}
\end{equation}
\begin{equation}
b= \frac{1}{2^{10} 7 \pi^2}(1744+171 \pi^2 - 3276\, \gamma) = 0.0218,
\label{3.9b}
\end{equation}
\begin{equation}
c = \frac{120}{7} =17.1429,  
\label{3.9c}
\end{equation}
and
\begin{equation}
d = \frac{1728}{7}=246.86\; . 
\label{3.9d}
\end{equation}
From (\ref{3.7}) we have
\begin{eqnarray}
0 &=& \mu^{1-2\epsilon} \frac{d \lambda_0}{d\mu} 
\nonumber \\
&=&2 \epsilon \left( \lambda + \frac{\lambda_1}{\epsilon} + \cdots \right)
+ \left( \mu \frac{\partial \lambda}{\partial\mu} + 
 \mu \frac{\partial \lambda}{\partial\mu} 
\frac{\partial \lambda_1}{\partial\lambda}\frac{1}{\epsilon} +
\mu \frac{\partial \alpha}{\partial\mu} 
\frac{\partial \lambda_1}{\partial\alpha}\frac{1}{\epsilon}
+ \cdots \right)\; ,
\label{3.10}
\end{eqnarray}
and using (\ref{3.3}) we get
\begin{eqnarray}
\beta_{\lambda} &=& \mu \frac{\partial \lambda}{\partial \mu}
\nonumber \\
&=& \left(\alpha \frac{\partial }{\partial \alpha}
+ 2 \lambda \frac{\partial }{\partial \lambda} - 2 \right) \lambda_1
(\alpha,\lambda) - 2 \lambda \epsilon
\nonumber \\
&=& 2 a (\lambda^2 - c \lambda \alpha^2 + d \alpha^4 - 2 b \lambda^3 )         \;\;\;(\mbox{for} \;\;\epsilon\rightarrow 0) \; 
\label{3.11}
\end{eqnarray}

Up to 2 loops ( terms in $\lambda^2$, $\lambda \alpha^2$ and
$\alpha^4$ ) this result qualitatively coincides with that of \cite{4}
 for this same model.  It does not, however, coincide with the result
of \cite{15} ( we will discuss this fact in the conclusions ). As can
be seen from (\ref{3.9c}), the contribution of the 4 loops graphs (
term in $\lambda^3$ ) is small and will not qualitatively change the
results for $\beta_{\lambda}$.
 
Making $\alpha =0$ we go to the pure $(\phi^\dagger \phi)^3$ model. In
this case $\beta$ starts at zero for $\lambda = 0$ and increases
monotonically with $\lambda$ \cite{1}. The model presents an infrared
(IR) fix point at $\lambda=0$.

For $\alpha \neq 0$ a drastic change occurs. In this case $\beta$
starts at $(4\,a\,d\,\alpha^4) > 0$ for $\lambda=0$ and never vanishes
in the perturbative range of the two coupling constants.  A similar
behavior of the $\beta$ function, already in one loop order, is shown
in the Coleman-Weinberg model (CW) \cite{15a} in (3+1) dimensions.
There, a dynamical symmetry breakdown occurs and masses are generated
for the two fields.  In \cite{15} the effective potential was
calculated in two loops and a breakdown of symmetry was also shown to
appear. We would like to stress that our results for $\Gamma^2$ and
$\Gamma^6$ are compatibles with that conclusion.  The $\Gamma^2(v)$
for the displaced field $\psi=\phi-v$, were $v$ is a constant with
dimension $(mass)^{1/2}$, would be written, in terms of the functions
that we calculated for $\phi$, as a series of the form $\Gamma^2(v) =
\Gamma^2 +( v^2/2) \Gamma^4 + (v^4/4!) \Gamma^6 + \cdots $ As can be
seen from the graphs proportional to $\alpha^4$ in figure 5,
$\Gamma^6$ receives a constant ( independent of p ) finite
contribution.  As consequence, $\Gamma^2(v)$ will have a singularity
displaced to some non null value of $p^2$, compatible with a non null
dynamically generated mass for $\phi$.

The anomalous dimensions of the fields  $A_\mu$ and $\phi$ are given by
\begin{equation}
\gamma_A = \frac{1}{2} \frac{\mu}{Z_A} \frac{dZ_A}{d\mu} 
\label{3.13}
\end{equation}
\begin{equation}
\gamma_\phi = \frac{1}{2} \frac{\mu}{Z_\phi} \frac{dZ_\phi}{d\mu} \; .
\label{3.14}
\end{equation}
As shown in section II, $Z_A= 1+ B =1$ and so $\gamma_A =0$. By
writing
\begin{equation}
Z_\phi = 1 + A = 1 + \frac{a_1 (\alpha,\lambda)}{\epsilon} +...
\label{3.15}
\end{equation}
where $a_1$ is given in (\ref{2.9}) we can write (\ref{3.14}) in the
form
\begin{equation}
2 \left( 1+ \frac{a_1}{\epsilon} + ...\right) \gamma_\phi = 
\mu \frac{\partial\lambda}{\partial\mu} \frac{\partial a_1}{\partial\lambda}
\frac{1}{\epsilon} + \mu\frac{\partial\alpha}{\partial\mu} 
\frac{\partial a_1}{\partial\alpha}\frac{1}{\epsilon} 
+ ... \; ,
\label{3.16}
\end{equation}
and using (\ref{3.3}) and (\ref{3.11}) we get

\begin{equation}
\gamma_\phi = -\lambda\frac{\partial a_1}{\partial \lambda}
-  \frac{\alpha}{2} \frac{\partial a_1}{\partial \alpha}\; .
\label{3.17}
\end{equation}
By substituting $a_1$, from (\ref{2.9}), in (\ref{3.17}) we have
\begin{equation}
\gamma_\phi = -\frac{7}{48\pi^2} \alpha^2 + \frac{1}{3^2.2^{12} \pi^4}
\lambda^2 \; .
\label{3.18}
\end{equation}

The contribution in $\alpha^2$ qualitatively agrees with the result of
\cite{4}. The term in $\lambda^2$ comes from 4 loops graphs ( not
calculated in \cite{4} ) and is very small compared to the term in 
$\alpha^2$.  It can be seen from (\ref{3.17}), that the scalar
field dimension, $D_\phi =\frac{1}{2} + \gamma_\phi$, decreases with
the coupling to the CS field. As it is well known, in non perturbative
approach in quantum mechanics, the coupling of matter fields to a CS
field, changes the spin and statistics of the matter fields, driving
bosons into anyons and also, for strong enough coupling, into
fermions.  Based on these results, there is a conjecture in the
literature\cite{11} that, even in perturbative quantum field approach
(in which the strength $\alpha \ll 1$) the dimension of a boson
coupled to a CS should receive an increase in the direction of the
fermion dimension $d_\psi =1$ ( for the corresponding problem of
fermions a decrease in the direction of the boson dimension should be
expected ). As shown in (\ref{3.18}) this conjecture is not realized:
the coupling to the CS field works in the direction of decreasing the
dimension of $\phi$.

To get a bit farther in testing this conjecture, we have also looked
at the anomalous dimensions of the composite operators
$[(\phi^\dagger\phi)^n]$, where n is an integer number.  As we are
mainly interested in the effect of the coupling of the boson to the CS
field, to simplify the calculations, we will restrict the
analyze to $\lambda=0$. In terms of monomials of
$\phi$ this composite operator can be written\cite{13}
\begin{equation}
[(\phi^\dagger \phi)^n]= Z_n (\phi^\dagger \phi)^n + 
Z_{n-1}^0 (\phi^\dagger \phi)^{n-1} + Z_{n-2}^2 (\phi^\dagger \phi)^{n-2}
(\phi^\dagger \partial^2 \phi) + ...\; .
\label{3.19}
\end{equation}
Determination of the $Z_m^{i}$ ($m \leq n$) require the calculation of the
divergent parts of the 2m scalar field 1PI vertex functions with the
insertion of one integrated composite operator

\begin{equation}
\Gamma_{[(\phi^*\phi)^n]}(x_1,...,y_m)= \int d^3 z 
< T [(\phi^\dagger \phi)^n] (z) 
\phi (x_1) ... \phi(x_m) \phi^\dagger (y_1) ... \phi^\dagger (y_m) > 
\;  ,
\label{3.20}
\end{equation}
or, in momentum space, the $\Gamma_{[(\phi^*\phi)^n]}(p_1,...,p_{2m})$
function with zero momentum $q$ entering through the special vertex
$[(\phi^\dagger \phi)^n]$. Up to order $\alpha^2$, the divergent
graphs contributing to $\Gamma_{[(\phi^*\phi)^n]}(p_1,...,p_{2n})$ are
shown in figure 6. In figure 7 we draw some of the graphs that could
contribute to $\Gamma_{[(\phi^*\phi)^n]}(p_1,...p_{2(n-1)})$.
Diagrams in figure 7, are in fact all nulls, what imply that the
renormalization parameters $Z^{i}_{n-1}$ also vanish.  The same can be
shown to be true for all $Z^{i}_m$ which any $m<n$. So, the right side of
(\ref{3.20}) reduces to only the first monomial and $[(\phi^\dagger
\phi)^n]$ does not mix with other operators ( mixing will however
appear if we consider $\lambda \neq 0$ ). Its renormalization only
requires the calculation of $Z_n$, what means to calculate the
divergent parts of the graphs in figure 6. The involved Feynman
integrals are the ${\cal G} (p,q)$ and ${\cal H} (p)$ from figure 5.
By writing $Z_n = 1+ A_n$ we have

\begin{equation}
\Gamma_{[(\phi^*\phi)^n]}(p_1,...,p_{2n})= (n!)^2 [A_n - (4 n^2 -2n) 
\alpha^2 {\cal G} - 2 i n^2 \alpha^2 {\cal H} ] + \mbox{finite graphs}\; ,
\label{3.21}
\end{equation} 
and we have for $A_n$
\begin{eqnarray}
A_n &=&  \mbox{DivPart}\;\{ (4n^2 -2 n)\alpha^2 {\cal G} 
+2 i n^2 \alpha^2 {\cal H}\} 
\nonumber \\
&=& - \frac{4n^2-n}{16 \pi^2} \frac{\alpha^2}{\epsilon} \; ,
\label{3.22}
\end{eqnarray}
where "DivPart" stands for keeping only the divergent part of the 
following expression.

With these results for $Zî_m$ and (\ref{2.9}) for $Z_\phi$, the
equation (\ref{3.19}) rewritten in terms of the unrenormalized (see
also (\ref{2.2a})) field $\phi_0$, becomes
\begin{equation}
[(\phi^\dagger \phi)^n]= Z^{-1}_{cn} (\phi^\dagger_0 \phi_0)^n \; ,
\label{3.23}
\end{equation}
where
\begin{equation}
 Z_{cn}= ({Z}_n)^{-1} (Z_\phi)^n = 1 + \frac{a_{cn} 
(\alpha)}{\epsilon}+ ... \; ,
\label{3.24}
\end{equation}
and
\begin{equation}
 a_{cn} (\alpha)= \frac{\alpha^2}{4 \pi^2} \left( n^2 + \frac{n}{3}
 \right) \; .
\label{3.25}
\end{equation}

By deriving the two sides of (\ref{3.23}) with respect to $\mu$ and
remembering that $\phi_0$ is independent of $\mu$ we have
\begin{equation}
\mu\frac{d}{d\mu} [(\phi^\dagger \phi)^n] = 
- \gamma_{cn} [(\phi^\dagger \phi)^n]\; ,
\label{3.26}
\end{equation}
where
\begin{equation}
 \gamma_{cn} = \frac{\mu}{Z_{cn}} \frac{d Z_{cn}}{d\mu} \;,
\label{3.27}
\end{equation}
is the anomalous dimension of the composite operator. Going through the same
steps that leads (\ref{3.15}) to (\ref{3.17}) we get
\begin{equation}
\gamma_{cn} = -\alpha \frac{d a_{cn}}{d\alpha} 
= -\frac{\alpha^2}{2\pi^2} \left(n^2 + \frac{n}{3}\right) \; .
\label{3.28}
\end{equation}

The dimension of the composite operator 
$ [(\phi^\dagger \phi)^n]$ becomes
\begin{equation}
D_{   [(\phi^\dagger \phi)^n]} = n 
 -\frac{\alpha^2}{2\pi^2} \left(n^2 + \frac{n}{3}\right) \; .
\label{3.29}
\end{equation}

This result is in disagreement with \cite{6}. Their calculation seems
to miss the contribution of the second graph in our figure 6. But it
is not this fact what makes the major difference. Our counting of the
combinatorial factors of the graphs in figure 6, gives a term
proportional to $n^2$ ( besides the term in $n$ ), different from
theirs which is only proportional to $n$.

No matter if the composite operator is super-renormalizable ($n<3$),
renormalizable ($n=3$) or non-renormalizable ($n>3$), the effect of
the coupling to the CS field is to lower its dimension.  Nevertheless,
the lowest non-renormalizable operator, $(\phi^\dagger \phi)^4$, with
effective dimension: $D_4 = 4 - \frac{52}{6\pi^2} \alpha^2$ will
never, in the perturbative regime, be driven to be renormalizable.
Yet, due to the quadratic dependence of the anomalous dimension on
$n$, given any $\alpha \ll 1$, the operators $[(\phi^\dagger \phi)^n]$
with $n$ bigger than $n_c \simeq \frac{2\pi^2}{\alpha^2} -\frac{10}{3}
\gg 1$ have their operator dimensions driven to values smaller than 3.

To finish this section, let us look at the renormalization group equations 
for the $\Gamma_{(2n)}(p,\lambda,\alpha,\mu)$ functions ( p is a short 
for the 2n external momenta ). As the 4 loops contributions are very small 
we will restrict the analyses to 2 loops. As $\beta_{\alpha}$ and 
$\gamma_{A}$ are null we have the renormalization group equation 
\begin{equation}
(\,\, \mu \frac{\partial}{\partial \mu} + \beta_{\lambda} \frac {\partial}
{\partial \lambda} - 2n \gamma_{\phi} \,)\,\, \Gamma_{(2n)} ( p,\lambda,\alpha,\mu ) = 0.
\label{3.30}
\end{equation}
The solution of this equation can be written as 
\begin{equation}
\Gamma_{(2n)} (\, p,\lambda,\alpha,\mu ) =\Gamma_{(2n)} 
(\, p,\bar{\lambda},\alpha,\mu s_{\bar{\lambda},\lambda})\,\,  
\,\, s_{\bar{\lambda},\lambda}^{ \,\,\,\,2n \gamma_{\phi}} 
\label{3.31}
\end{equation}
where we used the fact that up to two loops, $\gamma_\phi = -\frac{7}{48\pi^2}
\alpha^2$ does not change with s.  In the above equation,
$s_{\bar{\lambda}\lambda}$ stands for the solution of
\begin{eqnarray}
s \frac{d}{d s} \bar{\lambda}&=&\beta_{\lambda} 
(\bar{\lambda})\nonumber\\
&=& 2 a (  \bar{\lambda}^2 - c  \alpha^2 \bar{\lambda} +d \alpha^4 ),
\label{3.32}
\end{eqnarray}
with the condition $\bar{\lambda}(s=1) = \lambda$, that is :
\begin{eqnarray}
s_{\bar{\lambda}\lambda} &=& exp\left(\frac {1}{a f \alpha^2} 
\left[\,\, tan^{-1} \left( \frac {2 \bar{\lambda}}{f \alpha^2} - 
\frac{c}{f} \right)- tan^{-1} \left( \frac {2 \lambda}
{f \alpha^2} - \frac{c}{f} \right)\,\, \right] \right) \nonumber\\              &\cong& exp\left(\frac {2.86}{\alpha^2} \left[\,\, tan^{-1} 
\left( \frac {\bar{\lambda}}
{12 \alpha^2} - 0.71 \right)- 
tan^{-1} \left( \frac {\lambda}
{12 \alpha^2} - 0.71 \right)\,\, \right] \right), 
\label{3.33}
\end{eqnarray}
where $f = ( 4 d - c^2 )^{1/2}$. As $\beta_{\lambda}$ is non null for $\lambda=0$ ( for $\alpha\neq 0$ ) 
this equation is well defined if we choose $\lambda = 0 $. With this 
choice in (\ref{3.33}) we can write 
\begin{equation}
\Gamma_{(2n)} (\, p,\bar{\lambda},\alpha,\mu ) =\Gamma_{(2n)} 
(\, p,0,\alpha,\mu\,s_{\bar{\lambda}\,0}^{\,\,-1}                       )\,\,  
\,\, s_{\bar{\lambda}\,0}^{\,\,\,\,-2n \gamma_{\phi}}.
\label{3.34}
\end{equation}
This equation shows that, up to two loops, the $\Gamma_{(2n)}$
functions of the model defined by Lagrangian (\ref{2.3}), can be get
from the corresponding $\Gamma_{(2n)}$ for the model where only the
interaction term with the $A_{\mu}$ field is present, or what is equivalent,
from the calculation of the sub set of diagrams contributing to
$\Gamma_{(2n)}$, which only involves the interaction vertex with the
$A_{\mu}$ field.  A short inspection of the CW \cite{15a} results, shows that a
similar fact is also true for that model ( at least in one loop ).
\section{ Conclusions} 
The coupling to the CS field lowers the dimension of $\phi$ and of
$(\phi^\dagger \phi)^n$ . This goes in the opposite direction of the
conjecture that the transmutation of the boson into anyon ( due to the
coupling to the CS field ) should be signaled by the dimension of
these operators to increase in the direction of the canonical
dimension of a fermion field $\psi$ and their composite operators
$(\psi^\dagger \psi)^n$, respectively.

In the present paper, as in previous calculations in the literature,
the function $\beta_{\alpha}$ and the anomalous dimension of the CS
field are shown to vanish; the CS coupling constant $\alpha$ does not
run with the change of the energy scale.  The function
$\beta_{\lambda}$ instead, shows a drastic change in the presence of
the CS field. From an IR trivial fix point for the pure $\lambda
(\phi^* \phi)^3$ interaction, the model is driven, to a phase in which
no fix point appears for $\beta_{\lambda}$, in a behavior similar to
that of $\beta_{\lambda}$ for the model of Coleman-Weinberg \cite{15a}.

In \cite{15}, the renormalization group functions were
calculated up to 2 loops, although their main aim was to study the
effective potential and dynamical symmetry breakdown. The model of
\cite{15}, defined by their Lagrangian (2.1) can be made to coincide
with ours by deleting their $\lambda (\phi^* \phi)^2$ interaction and
the $m^2 (\phi^* \phi)$ mass term, that is, by making their $\lambda$
and $m$ zero. Considering also, that their coefficient, $\nu$, of the
$(\phi^* \phi)^3$ interaction, differs from our $\lambda$ by a factor
of 2/5, what also implies in a 2/5 factor of difference in the
corresponding $\beta$ functions, their results ( equations (10.7-9)
and (11.8) ), after translated to our notation, can be summarized as:
$\bf{1.}$ $\beta_{\alpha}=0$ and $\gamma_A=0$.  These results are in
agreement with our equation (\ref{3.4}) and the observation below
equation (\ref{3.14}). $\bf{2.}$ $\gamma_{\phi}={\cal
  O}(\lambda^{2})$, and $\beta_{\lambda}= 2a\lambda^2 +{\cal
  O}(\lambda^{3})$, both independent of $\alpha$.  Our results
(\ref{3.11}) and (\ref{3.18}) differ from these last ones by terms
dependents on the CS coupling $\alpha$.  Their conclusion is that the
model has an IR trivial fix point in $\lambda$. Ours instead, is that
$\beta_{\lambda}$ never vanishes, a result similar to that of CW in a
model in which a dynamical symmetry breakdown occurs. A dynamical
symmetry breakdown was also seen in \cite{15} for the present model. Our
result for $\beta$ looks so, in accordance with their result on
symmetry breakdown.

The discrepancies between ours and the $\beta$ function of \cite{15}
can be attributed to the different regularization schemes we are
using.  In \cite{15}, the model is regularized through a full
dimensional regularization, by extending out of 3D, all the tensor
structures ( including the definition of the $\epsilon_{\mu \nu \rho}$
) that appear in the Feynman graphs. As they conclude, in that method,
the renormalizability of the model is only achieved, if an extra
regularization, represented by a Maxwell term for the $A^\mu$ field (
besides the CS one ), is introduced. Their method requires, that this
extra regularization be dismissed ( their parameter ``a'' taken to
zero ), only after the continuation back to 3D is made. As can be seen
from their results (11.8), some of their $\beta$ functions become
singular, when $a \rightarrow 0$, showing that a better understanding
of the structure of the renormalization group equation is still
lacking in that method. Also, as discussed in their Section 10, if a
regularization directly in 3D ( exists and ) were used,
$\gamma_{\phi}$ and $\beta_{\lambda}$ would be expected to depend also
on $\alpha$.

In this paper we used the Dimensional Reduction regularization scheme,
in which all the tensor contractions are first made in 3D and only the
remaining scalar Feynman integrals are extended out of 3D. We
explicitly verified that this method controls all the UV infinities
and preserves the Ward identities ( and so, the gauge covariance ) up
to the order of approximation in which we are working ( 2 loops in
graphs involving the CS propagator and 4 loops in graphs only
involving the scalar propagator ). Although we can not say that it is
a regularization directly in 3D, our results is consistent with the 
above mentioned discussion in \cite{15}.

As a definitive answer to this problem is desirable, we are presently
working in a related model, using a direct in 3D version of the BPHZ
renormalization method.  The preliminary results confirm those 
of the present paper for the renormalization group functions, together
with the dynamical symmetry breakdown got in \cite{15}.

To finalize we would like to summarize the results of two previous
papers \cite{12}, in which we studied the scale behavior of fermions
interacting with a CS field.  In the first one, a single fermion with
its most general 4-fermion ( non renormalizable ) self interaction
$g(\bar{\psi}\psi)^2$ was considered.  We saw that, although $\psi$
gets a negative anomalous dimension, making its operator dimension to
approach that of a boson, no definite pattern of approach to a bosonic
scale behavior due to the interaction with the CS field is seen for
composite operators: the super-renormalizable operator
$\bar{\psi}\psi$ gets a negative anomalous dimension, but the
non-renormalizable operator $(\bar{\psi}\psi)^2$ gets a positive one.
In the second paper an extended version of this model with N ( small )
fermion fields, with their most general 4-fermion interaction:
$g(\bar{\psi}\psi)^2 + h(\bar{\psi}\gamma^{\mu}\psi)^2$ was
considered.  We studied operators of canonical dimension four.
We showed that one of them has positive anomalous dimension, other
has a very small negative anomalous dimension and the third one, more
interesting from the renormalization view point, has a negative
anomalous dimension, making, through a fine tuning of the coupling
constants, its operator dimension as close to 3 as wanted .
Nevertheless, no general pattern of approach to a bosonic like
behavior ( negative anomalous dimension ), as advanced by the
conjecture in the literature, was seen.

\appendix
\section{The Ward Identities}

The two relations among the counterterms: A to E can be get from the
WI among the 1PI 4-linear photon-scalar vertex, $\Gamma_{\mu\nu}$, the
trilinear photon-scalar vertex, $\Gamma_{\mu}$, and the scalar self
energy $\Gamma_2$. In tree approximation they are given by ( see
figure 1 ): $\Gamma_{\mu\nu}=2 i e^2 {\mu}^{\epsilon} g_{\mu\nu}$,
$\Gamma_{\mu}=-i e {\mu}^{\epsilon/2}(p\prime+ p)_{\mu}$ and
$\Gamma_2=i A p^2$. It is ease to use that they satisfy the relations
\begin{equation}
q^{\mu} \Gamma_{\mu} (q;p,p^\prime) =-e \mu^{\frac{\epsilon}{2}}    
\left\{ \Gamma_2(p^\prime) -\Gamma_2 (p) \right\}\;,
\label{A1}
\end{equation}

\begin{equation}
q^\mu \Gamma_{\mu\nu} (q,q^\prime;p,p^\prime) =
-e \mu^{\frac{\epsilon}{2}} 
\left\{ \Gamma_\nu (q^\prime; p^\prime -
q^\prime, p^\prime) - \Gamma_\nu (q^\prime; p, p+q^\prime ) \right\}
\; .
\label{A2}
\end{equation}

As we explicitly verified these relations are, in fact, valid up to 2-loop
order. Instead of considering the WI among the sum of all graphs up to
2-loops contributing to each of the 3 vertex functions above, we can
take advantage of the fact that they can be separated in sub classes
to be seen to be separately related through the the WI (\ref{A1}) and
(\ref{A2}). As an example consider the graphs (8-a) to (8-h)
contributing to $\Gamma_\mu$ and (9-a) and (9-b) contributing to
$\Gamma_2$. Let us call $\tilde{\Gamma}_\mu$ the sum of contributions of
diagrams (8-a) to (8-q) and $\tilde{\Gamma_2}$ the graph (9-a). Let
also $\tilde{D}$ and $\tilde{A}$ be the possible divergent
contributions to the counterterms $D$ and $A$, chosen so as to
make finite the sums of graphs in figure 8 and 9, respectively. By
using dimensional reduction regularization, and explicitly
writing all the Feynman integrands involved, we can verify that
\begin{equation}
q^\mu \left\{ \tilde{\Gamma}_\mu (q;p,p^\prime) -
ie \mu^{\frac{\epsilon}{2}} (p^\prime +p)_\mu \tilde{D}
\right\} = -e  \mu^{\frac{\epsilon}{2}} 
 \left\{( \tilde{\Gamma_2} (p^\prime)+i p^{\prime 2} \tilde{D})-
 (\tilde{\Gamma_2} (p)+i p^{ 2} \tilde{D})\right\}\; .
\label{A3}
\end{equation}
As $\tilde{D}$ is chosen so as to make the bracket in the left side of
these equations finite, the right side is also finite, what implies
that: $ip^2 \tilde{D} = -\mbox{DivPart}\{\tilde{\Gamma_2}(p)\}\equiv
ip^2 \tilde{A}$, that is $ \tilde{D}=\tilde{A}$. A more direct
verification is obtained by explicitly calculating:
\begin{equation}
ie  \mu^{\frac{\epsilon}{2}} (p^\prime +p)_\mu \tilde{D} =
\mbox{DivPar} \{ \tilde{\Gamma}_\mu (q;q,p^\prime)\}
\label{A4}
\end{equation}
and
\begin{equation}
ip^2 \tilde A = -\mbox{DivPart} \{ \tilde{\Gamma_2} (p)\}\; .
\label{A5}
\end{equation}
The only really divergent graphs contributing to $\tilde{\Gamma}_\mu$
are (8-a) and (8-g). By going through the calculation of the divergent
parts of (8-a) plus (8-g) as exemplified in Appendix B we get
\begin{eqnarray}
ie (p^\prime +p)_\mu \tilde{D} &=&
\mbox{DivPart}\left\{ (-ie)^3 (ie^2) (i)^3\frac{12}{3!}
\int {\cal D}k  \int {\cal D}k^\prime 
\varepsilon_{\beta\nu\rho} \frac{k^\rho}{k^2}
\varepsilon_{\alpha\mu\gamma} \frac{k^{\prime\gamma}}{k^{\prime 2}}
\right.\nonumber \\
& &\left. \times\frac{ (2p+k)_\beta (2p + 2k +k^\prime)_\alpha 
(2p+2k^\prime +k)_\nu}{ (p+k)^2 (p+k^\prime)^2 (p+k+k^\prime)^2} 
+p\leftrightarrow p^\prime \right\}
\nonumber \\
&=& i \frac{e^5}{12 \pi^2} \frac{1}{\epsilon} p_\mu + 
  i \frac{e^5}{12 \pi^2} \frac{1}{\epsilon} p^\prime_\mu
\label{A6}
\end{eqnarray}
that is: $ \tilde{D} = \frac{\alpha^2}{12 \pi^2} \frac{1}{\epsilon}$.

For $\tilde{A}$ we have 
\begin{eqnarray}
ip^2 \tilde{A} &=& -\mbox{DivPart}\{ \tilde{\Gamma} (p)\} \nonumber \\
&=& -\mbox{DivPart}\left\{\mbox{Graph (3-b)}\right\}\nonumber \\
&=& i\frac{e^4}{12 \pi^2} \frac{1}{\epsilon} p^2
\label{A7}
\end{eqnarray}
as given by  (2.8). So, we have $\tilde{D}=\tilde{A} = 
\frac{\alpha^2}{12 \pi^2} \frac{1}{\epsilon}$.

An example of subset of graphs that match through the $2^d$ WI are
depicted in Fig. 10 and Fig. 11. The identification of
$\tilde{E}^\prime = \tilde{D}^\prime$ follows through the same steps
as in the example above.
%%%%%%%%%%%%%%%%%%%%%%%%%%%%%%%%%%%%%%%%%%%%%%%%%%%%%%%%%%%%%%%%%%%%
\section{Feynman Integrals}

To illustrate the method adopted to get the divergent parts of the
Feynman integrals that appear in the paper, we will explicitly show as
examples the calculation of the diagrams 5.a, 5.b, 5.d and 5.f.  Let
us start with (5.a). In the figure, $\Delta (k)=i/(k^2+i\eta)$ as
usual, and $\Delta_2 (k)$ and $\Delta_3 (k)$ stand for the subgraphs
formed respectively by 2 and 3 scalar propagators connecting 2
vertices, with total momentum $k$ passing through. Its integration can
be done successively one loop at time, first getting $\Delta_2$ and
then $\Delta_3$.  $\Delta_2 (p)$ is given by
\begin{equation}
\Delta_2 (p) =\int {\cal D} k \frac{i}{k^2 +i\eta} \frac{i}{(k+p)^2
  +i\eta}\; ,
\label{B1}
\end{equation}
where ${\cal D}k = \mu^\epsilon d^{3-\epsilon} k
/(2\pi)^{3-\epsilon}$. By introducing a Feynman parameter through the
use of the identity
\begin{equation}
\frac{1}{A^\alpha B^\beta}= \frac{\Gamma (\alpha+\beta)}{\Gamma
(\alpha) \Gamma (\beta)} \int_0^1 dx
\frac{x^{\alpha-1} (1-x)^{\beta-1}}{[Ax + B(1-x)]^{\alpha+\beta}}\;,
\label{B2}
\end{equation}
the $k$ integration can be done\cite{13} and then
the parametric integration\cite{14} to give
\begin{equation}
\Delta_2 (p) = -i \frac{(4\pi
  \mu^2)^{\frac{\epsilon}{2}}}{(4\pi)^{\frac{3}{2}}} 
\frac{\Gamma^2 \left(\frac{1}{2} -\frac{\epsilon}{2}\right)
\Gamma \left(\frac{1}{2} +\frac{\epsilon}{2}\right)}
{\Gamma (1-\epsilon)} (-p^2 -i\eta)^{- \left(\frac{1}{2}
  -\frac{\epsilon}{2}\right)} \;.
\label{B3}
\end{equation}
$\Delta_3 (q)$ can be written as
\begin{equation}
\Delta_3 (q) = \int {\cal D} p \frac{i}{(p+q)^2+i \eta} \Delta_2 (p)\;
. 
\label{B4}
\end{equation}

This integration can also be done following the same steps as for
$\Delta_2 (p)$, after explicitly substituting in this last equation,
the expression (\ref{B3}) for $\Delta_2 (p)$. The result is
\begin{equation}
\Delta_3 (q) = -\frac{i}{(4\pi)^3} 
\frac{\Gamma^3 \left(\frac{1}{2} -\frac{\epsilon}{2}\right)
\Gamma \left(\epsilon\right)}
{\Gamma \left(\frac{3}{2} -\frac{3\epsilon}{2}\right)}
\left( -\frac{4\pi \mu^2}{q^2 +i\eta}\right)^\epsilon \; .
\label{B5}
\end{equation}
For 5.d we have
\begin{equation}
{\cal W} (q,p) = \int {\cal D} k \frac{i}{(k+q)^2 +i\eta}
\Delta_2 (k+p) \Delta_3 (k) \; ,
\label{B6}
\end{equation}
This integral can be done by first reducing the 3 denominators to a
single one, by twice using (\ref{B2}) to get a single denominator and
then doing the $k$ integration\cite{13}. In terms of the 2 remaining
Feynman parameters it has the form
\begin{equation}
{\cal W} (q,p) = -\frac{1}{(4\pi)^6} \frac{
\Gamma \left(2\epsilon\right)
\Gamma^5 \left(\frac{1}{2} -\frac{\epsilon}{2}\right)}
{\Gamma \left(1 -\epsilon\right)
\Gamma \left(\frac{3}{2} -\frac{3\epsilon}{2}\right)}
\left( -\frac{4\pi \mu^2}{p^2 +i\eta}\right)^{2\epsilon} I_\epsilon (q,p)\; ,
\label{B7}
\end{equation}
where $I_\epsilon (q,p)$ is given by
\begin{equation} 
I_\epsilon (q,p)= \int_0^1 dy (1-y)^{\epsilon -1} f_\epsilon (y)\; ,
\label{B8}
\end{equation}
and
\begin{equation} 
f_\epsilon (y)= \int_0^1 dx x^{-\frac{1}{2} +\frac{\epsilon}{2}}
y^{\frac{1}{2} +\frac{\epsilon}{2}}
\left\{ \frac{q^2}{p^2} [y^2 (1-x)^2 - y (1-x)]+
2\frac{p.q}{p^2} y^2 x (1-x) + y x (yx -1)\right\}^{-2\epsilon}\; .
\label{B8b}
\end{equation}

$I_\epsilon (q,p)$ has a single pole in $\epsilon$ coming from the
integration region in the vicinity of $y=1$. As (\ref{B7}) already has
a factor $\Gamma(2\epsilon)$ the integral (\ref{B6}) will present both a
single and a double pole in $\epsilon$. To separate their
contributions we must calculate the first 2 terms (single pole and the
$\epsilon$ independent term) of the Laurent expansion of $I_\epsilon
(q,p)$.  We have
\begin{eqnarray}
I_\epsilon (q,p) &=& I_{1\epsilon} (q,p)+  I_{2\epsilon}
(q,p)\nonumber \\
&=& \frac{A_1}{\epsilon} + (B_1 + B_2) + (C_1+C_2) \epsilon + ...\; ,
\label{B9}
\end{eqnarray}
where
\begin{eqnarray}
 I_{1\epsilon} (q,p)&=& \int_0^1 dy (1-y)^{\epsilon -1} f_\epsilon (1)
\nonumber \\
&=&\frac{A_1}{\epsilon} + B_1  + C_1 \epsilon + ...\; , 
\label{B10}
\end{eqnarray}
\begin{eqnarray}
 I_{2\epsilon} (q,p)&=& \int_0^1 dy (1-y)^{\epsilon -1} 
(f_\epsilon (y) - f_\epsilon (1)
\nonumber \\
&=& B_2  + C_2 \epsilon + ...\; , 
\label{B11}
\end{eqnarray}
where $A_1$, $B_1$ and $B_2$ are still to be determined. $B_2$ is given by
\begin{equation}
B_2=I_{20} (q,p) = \int_0^1 dy (1-y)^{-1} \int_0^1 dx (y^{\frac12}-1)
x^{-\frac12}= 4(-1+\log 2)\; .
\label{B12}
\end{equation}
$A_1$ and $B_1$ come from
\begin{eqnarray}
I_{1\epsilon} (p,q) &=& \int_0^1 dy  (1-y)^{\epsilon -1} 
\int_0^1 dx x^{\frac{\epsilon}{2} -\frac{1}{2}} (x^2-x)^{-2\epsilon} 
\left\{ \frac{(q-p)^2}{(p^2)}\right\}^{-2\epsilon} \nonumber \\
&=&(-1)^{-2\epsilon} \left(\frac{p^2}{(q-p)^2}\right)^{2\epsilon}
\frac{\Gamma \left(\frac{1}{2} -\frac{3\epsilon}{2}\right)
\Gamma \left(1-2\epsilon\right)\Gamma \left(\epsilon\right)}
{\Gamma \left(\frac{3}{2} -\frac{7\epsilon}{2}\right)
\Gamma \left(1+\epsilon\right)}\; .
\label{B13}
\end{eqnarray}
The results are $A_1=2$ and $B_1=2 \left\{ 5 -2\log 2
  -\frac{7\gamma}{2} + 2 \log
  \left(-\frac{p^2}{(p-q)^2}\right)\right\}$.  Multiplying the Laurent
expansion of $I_\epsilon (p,q)$ by the Laurent expansion of the
multiplying factor in (\ref{B7}) we get
\begin{equation}
{\cal W} (q,p) = -\frac{1}{2^{11} \pi^4} \frac{1}{\epsilon^2} -
\frac{1}{2^{10} \pi^4} \left\{ 4 -\frac{11}{4}\gamma + 
\log\left(-\frac{4\pi\mu^2}{(p-q)^2}\right)\right\} \frac{1}{\epsilon}
+ \mbox{finite part}.
\label{B14}
\end{equation}

Let us go to (5.b). The sub diagram $D_2(k)$ is given by
\begin{equation}
D_2(k)= \int {\cal D} q
\varepsilon^{\mu\nu\lambda}\frac{q_\lambda}{q^2+i\eta}
\varepsilon_{\nu\mu\rho}\frac{(q+k)^\rho}{(q+k)^2+i\eta}\; .
\label{B15}
\end{equation}
After contracting the tensors in ($2+1$) dimension we are left with
the (finite) integral
\begin{eqnarray}
D_2(k)&=& -2\int {\cal D} q \frac{k.q}{[q^2+i\eta][(q+k)^2+i\eta]}
\nonumber \\
&=& -\frac{i}{8}\frac{(4\pi\mu^2)^{\frac{\epsilon}{2}}}
{(-k^2-i\eta)^{-\frac{1}{2}+\frac{\epsilon}{2}}}\; ,
\label{B16}
\end{eqnarray} 
where $\epsilon$ was made zero whenever possible. Graph (5.b) is
given by
\begin{equation}
{\cal G} (p_1,p_2) =\int {\cal D} k \frac{i}{(p_1+k)^2 +i\eta} 
\frac{i}{(p_2+k)^2 +i\eta} D_2 (k) \; .
\label{B17}
\end{equation}
This integral is logarithmically divergent and their residue is
independent of $p_1$ and $p_2$. To get this residue it is sufficient to
calculate it for $p_2=-p_1$
\begin{equation}
{\cal G}\frac{}{}\mid_{p_2=-p_1} 
=\int {\cal D} k \frac{i}{[-(p+k)^2 -i\eta]^2}
\frac{1}{(-k^2-i\eta)^{-\frac{1}{2} +\frac{\epsilon}{2}}}\;,
\label{B18}
\end{equation}
where whenever possible we have put $\epsilon=0$. After introducing a
Feynman parameter through (\ref{B2}) and integrating in $k$ we get
\begin{equation} 
{\cal G}\frac{}{}\mid_{p_2=-p_1} = -\frac{1}{2^5 \pi^2} \Gamma
(\epsilon) (-p_1^2)^{-\epsilon} = -\frac{1}{32\pi^2}
\frac{1}{\epsilon} + ...\; .
\label{B19}
\end{equation}
Contribution of diagram (5.f) is given by
\begin{equation}
{\cal H}(p) = -i \int{\cal D}q {\cal D}k
\;\varepsilon^{\mu\nu\rho} \frac{(p+q)_\rho}{(p+q)^2}
\;\varepsilon^{\alpha\beta\lambda}\frac{q_\lambda}{q^2}
\;\frac{g_{\nu\alpha} (2k+p-q)_\mu (2k-q)_\beta}{(k+p)^2 (k-q)^2 k^2}
\; .
\label{B20}
\end{equation}
or
\begin{equation}
{\cal H} (p) =-2i
\varepsilon^{\mu\nu\rho}\varepsilon^{\alpha\beta\lambda}
g_{\nu\alpha}\frac{\mu^\epsilon}{(2\pi)^d}
\int {\cal D}q \frac{1}{(p+q)^2 q^2} I_{\beta\mu\lambda\rho}(q)\; ,
\label{B21}
\end{equation}
where
\begin{equation}
 I_{\beta\mu\lambda\rho}(q)=\int d^dk
\frac{2k_\beta k_\mu (q_\nu p_\rho + q_\nu q_\rho) +
k_\beta (q_\nu q_\rho p_\mu -q_\nu q_\mu p_\rho)}
{(k+p)^2 (k-q)^2 k^2}\; .
\label{B22}
\end{equation}
Using the identity
\begin{equation}
\frac{1}{ABC}= 2 \int_0^1 y dy \int_0^1 dx \frac{1}{[C(1-y)+ y (Ax +
  B(1-x))]^3}
\label{B23}
\end{equation}
and doing the $k$ integration  we get
\begin{eqnarray}
{\cal H} (p) &=& -\frac{2\; \Gamma \left(2-\frac{d}{2}\right) \mu^\epsilon}
{2^d \pi^{\frac{d}{2}}}
\;\varepsilon^{\mu\nu\rho}\;\varepsilon^{\alpha\beta\lambda}\; g_{\nu\alpha}
\int_0^1 y dy \int_0^1 dx \frac{1}{[-a']^{3-\frac{d}{2}}}
\int {\cal D} q \frac{1}{(p+q)^2 q^2}\nonumber\\ 
&\times&\left[ \frac{q_\nu q_\mu p_\rho p_\beta xy [(4-d) y(x-1) +1] +
q_\nu q_\rho p_\mu p_\beta x y [(4-d) xy -1]}
{\left[-q^2 - 2 q.p \frac{b'}{a'} -p^2
    \frac{c'}{a'}\right]^{3-\frac{d}{2}}}
\right.\nonumber \\
&&\left.\;\;\;\;\;\;\;\;\;\;\;\;\;\;\;+\;
 \frac{a'g_{\beta\mu} (q_\nu p_\rho +q_\nu q_\rho)}
{\left[-q^2 - 2 q.p \frac{b'}{a'} -p^2
    \frac{c'}{a'}\right]^{2-\frac{d}{2}}} \right]
\label{B24}
\end{eqnarray}
with $a'= y (x-1) [y(x-1)+1]$, $b'= x y^2 (x-1)$ and
$c'= x y (xy -1)$. The only divergent term is the last monomial in
the square bracket of (\ref{B24}). We can so, write
\begin{equation}
{\cal H} (p) = \frac{\; \Gamma \left(2-\frac{d}{2}\right) \mu^\epsilon}
{2^{d+1} \pi^{\frac{d}{2}}}
\;\varepsilon^{\mu\nu\rho}\;\varepsilon^{\alpha\beta\lambda}\; g_{\nu\alpha}
\int_0^1 dy y \int_0^1 dx \frac{1}{[-a']^{2-\frac{d}{2}}}
\frac{\mu^\epsilon}{(2\pi)^d} I_{DivPar} (p) + \mbox{fin parts}
\label{B25}
\end{equation}
where
\begin{equation}
I (p) = \int d^d q \frac{q_\nu q_\rho}
{(p+q)^2 q^2 \left[-q^2 - 2 q.p \frac{b'}{a'} -p^2
    \frac{c'}{a'}\right]^{2-\frac{d}{2}}}\; .
\label{B26}
\end{equation}
By reducing the denominators through the use of the identity
\begin{equation}
\frac{1}{A^\alpha B^\beta C^\gamma} =
\frac{\Gamma (\alpha+\beta
  +\gamma)}{\Gamma(\alpha)\Gamma(\beta)\Gamma(\gamma)} 
\int_0^1 dz \int_0^z dt \frac{t^{\gamma-1} (z-t)^{\beta-\alpha}}
{[A+ (B-A)z + (C-B) t]^{\alpha+\beta+\gamma}}\; .
\label{B27}
\end{equation}
and doing the integration in $q$ we get for the divergent
part
\begin{equation}
I_{DivPart} (p)= -i \frac{g_{\lambda\rho}}{2^{d+1}
  \pi^{\frac{d}{2}}[p^2]^\epsilon}
\frac{\Gamma(\epsilon)}{\Gamma \left(2-\frac{d}{2}\right)}
\int_0^1 dz \int_0^z dt t^{1-\frac{d}{2}} [a^{\prime\prime} -
b^{\prime\prime}]^{d-3}\; ,
\label{B28}
\end{equation}
where $a^{\prime\prime} = 1-z + \frac{b'}{a'} t$ and 
$b^{\prime\prime}=1-z + \frac{c'}{a'} t$.
By inserting (\ref{B28}) in (\ref{B26}), expanding in $\epsilon$ and
doing the parametric integrations we get
\begin{equation}
{\cal H} =\frac{i}{16\pi^2} \frac{1}{\epsilon} +\mbox{finite parts}\; .
\label{B30}
\end{equation}

%%%%%%%%%%%%%%%%%%%%%%%%%%%%%%%%%%%%%%%%%%%%%%%%%%%%%%%%%%%%%%%%%%%%
\center{ACKNOWLEDGEMENTS}

This  work was partially supported by the Brazilian agencies: Conselho 
Nacional de Desenvolvimento Cient\'{\i}fico e Tecnol\'ogico (CNPq), 
Coordenadoria de Aperfei\c coamento de Pessoal de Ensino Superior (CAPES)
and by Fundac\~ao de Amparo a Pesquisa do Estado de S\~ao Paulo (FAPESP).
\newpage
%%%%%%%%%%%%%%%%%%%%%%%%%%%%%%%%%%%%%%%%%%%%%%%%%%%%%%%%%%%%%%%%%%%%%%

\newpage
Figure Captions 

\begin{itemize}
\item
Figure 1. Feynman rules in the Landau gauge.
\item
Figure 2. Divergent diagrams contributing to the CS 
2 point function.
\item
Figure 3. Divergent diagrams contributing to the scalar 
field 2 point function.
\item

Figure 4. Divergent contributions to the scalar 
6 point function. Three others, not drawn, diagrams similar to 4.n, 
4.o and 4.p, 
but with the sense of all external lines reversed must also be 
considered.
\item

Figure 5. Representation of the  divergent integrals that appear 
in the diagrams of Figure 4.
\item

Figure 6. Divergent contributions to $\Gamma_{[(\phi^*\phi)^n]}(2n)$, that is, 
the 2n point function with one insertion of the composite 
operator $[(\phi^* \phi)^n]$.
\item

Figure 7. Some possible contributions to $\Gamma_{[(\phi^*\phi)^n]}(2(n-1))$ 
,the 2(n-1) point function with one insertion of the 
composite operator $[(\phi^* \phi)^n]$.
\item

Figure 8. An example of a family of diagrams contributing 
to $\Gamma_{\mu}(q;p,p')$.
\item

Figure 9. The diagrams contributing to $\Gamma(p)$, related by the 
Ward identity, to the family of diagrams in Figure 8.
\item

Figure 10. An example of family of diagrams contributing to 
$\Gamma_{\mu \nu}(q,q';p,p')$.
\item

Figure 11. The diagrams contributing to $\Gamma_{\mu}$, related 
by the Ward identity to the family of diagrams in Figure 10. 
\end{itemize}

\begin{thebibliography}{99}
\bibitem{1} C. A. Arag\~{a}o de Carvalho, Nucl. Phys. {\bf B119}, 401,
(1976); A. L. Lewis and F. W. Adams, Phys. Rev. {\bf B18}, 5099,
(1978); R. D. Pisarski, Phys. Rev. Lett. {\bf 48}, 574, (1982);
D. G. C. McKeon and G. Tsoupros, Phys. Rev. {\bf D46}, 1794, (1992);
{\bf 49}, 3065(E), (1994); J. R. Huish and D. Toms, Phys. Rev. {\bf
D49}, 6767, (1994); F. A. Dilkes, D. G. C. McKeon and K. Nguyen,
Phys. Rev. {\bf D57}, 1159, (1998).
\bibitem{2} F. Sch\"onfeld, Nucl. Phys. {\bf B185}, 157
  (1981);  S. Deser, R. Jackiw and S. Templeton,
 Phys. Rev. Lett. {\bf 48}, 975 (1982), Ann. Phys.(NY) {\bf 140}, 372 (1982). 
\bibitem{3} F. Wilczek, Phys. Rev. Lett. {\bf 48}, 1144 (1982);
  F. Wilczek, Phys. Rev. Lett. {\bf 49}, 957 (1982).
\bibitem{4}  L. V. Avdeev, G. V. Grigoryev, and D. I. Kazakov,
  Nucl. Phys. {\bf B382}, 561, (1992).
\bibitem{5}  G. W. Semenoff, P. Sodano, and Yong-Shi Wu,
  Phys. Rev. Lett. {\bf 62}, 715 (1989);
W. Chen, G, W. Semenoff and Y. S.  Wu, Mod. Phys. Lett.  {\bf A},
  Vol. 5, (1990); Phys. Rev. {\bf D44}, 1625, (1991) and
Phys. Rev. {\bf D46}, 5521, (1992).
\bibitem{6}  W. Chen and M. Li, Phys. Rev. Lett. {\bf 70}, 884
  (1993). 
\bibitem{8}C. G. Bollini and J. J. Giambiagi, Phys Lett {\bf
    40B}, 566 (1972); G. t'Hooft and M. Veltman, Nucl. Phys. {\bf
    B44}, 189 (1972) ; J. F. Ashmore, Lett. Nuovo Cimento {\bf 4}, 289
  (1972).
\bibitem{9} G. t'Hooft, Nucl. Phys. {\bf B61}, 455, (1993).
\bibitem{10} S. Coleman and B. Hill, Phys. Lett. {\bf 159B}, 184, (1985).
\bibitem{11} W. Chen, Nucl. Phys. {\bf B435}, 673, (1995).
\bibitem{12} V. S. Alves, M. Gomes, S. V. L. Pinheiro and A. J. da Silva, 
Phys. Rev. {\bf D59}, 045002 (1999) and Phys. Rev. {\bf D60}, 027701 (1999). 
\bibitem{13} J. Collins, {\it Renormalization}, Cambridge
  University Press, (1985).
\bibitem{14} I. S. Gradshteyn and I. M. Ryzhik, {\it Table of 
Integrals, Series and  Products}, Academic Press, (1963).
\bibitem{15} P. N. Tan, B. Tekin and Y. Hosotani, Nucl. Phys. 
{\bf B502}, 483, (1997), and Phys. Lett. {\bf B388}, 611, (1996).
\bibitem{15a} S. Coleman and E. Weinberg, Phys. Rev. {\bf D7}, 1888, (1975).
\end{thebibliography}
\end{document}